\documentclass[conference,letterpaper]{IEEEtran}

\usepackage{balance}
\usepackage{cite}
\usepackage{float}
\usepackage{multicol,multirow}
\usepackage{makecell,booktabs}
\usepackage{hyperref}
\hypersetup{    
    colorlinks=true,
    linkcolor=black,
    anchorcolor=black,
    citecolor=black,
    filecolor=black,
    menucolor=black,
    runcolor=black,
    urlcolor=black,
}

\usepackage{subcaption} 
\usepackage[utf8]{inputenc}

\usepackage{graphicx}
\usepackage{amsfonts}
\usepackage{amssymb}
\usepackage{amsmath,mathtools}
\usepackage{array}
\newcolumntype{P}[1]{>{\centering\arraybackslash}p{#1}}
\newcolumntype{L}[1]{>{\raggedright\arraybackslash}p{#1}}
\newcolumntype{R}[1]{>{\raggedleft\arraybackslash}p{#1}}
\usepackage{color}

\usepackage[labelsep=period]{caption}
\usepackage{bbm}
\usepackage{bm}
\usepackage{comment}
\usepackage{forest}
\usepackage{dblfloatfix}

\usepackage{etoolbox}

\newcommand{\jk}[1]{{\color{black}#1}}

\DeclareMathOperator*{\concat}{%
    \mathchoice%
        {\Big\Vert}%
        {\big\Vert}%
        {\Vert}%
        {\Vert}%
}

\title{\bf \Large
   Probabilistic Dynamic Line Rating Forecasting with Line Graph Convolutional LSTM
\vspace{-5mm}
}
\author{
    Minsoo Kim\textsuperscript{$\dagger$}, Vladimir Dvorkin\textsuperscript{$\ddagger$}, and Jip Kim\textsuperscript{$\dagger$}\\
    \textsuperscript{$\dagger$}Dept. of Energy Engineering,
    Korea Institute of Energy Technology\\
    \textsuperscript{$\ddagger$}Dept. of Electrical Engineering and Computer Science, University of Michigan, Ann Arbor\vspace{-4mm}
    \thanks{\vspace{-0mm}
    This work was supported by Basic Science Research Program through the National Research Foundation of Korea (NRF) funded by the Ministry of Education (No. RS-2024-00454017) and KENTECH Research Grant (202300008A).
    }    
    \vspace{-0mm}}

\begin{document}

\IEEEoverridecommandlockouts

\maketitle

\IEEEpubidadjcol

\begin{abstract}
    Dynamic line rating (DLR) is a promising solution to increase the \jk{utilization} of transmission lines by adjusting ratings based on real-time weather conditions. 
    Accurate DLR forecast at the \jk{scheduling} stage is thus necessary for system operators to proactively optimize power flows, manage congestion, and reduce the cost of grid operations. 
    However, the DLR forecast remains challenging due to weather uncertainty. 
    To reliably predict DLRs, we propose a new probabilistic forecasting model based on line graph convolutional LSTM. 
    Like standard LSTM networks, our model accounts for temporal correlations between DLRs across the planning horizon. The line graph-structured network additionally allows us to leverage the spatial correlations of DLR features across the grid to improve the quality of predictions. Simulation results on the synthetic Texas 123-bus system demonstrate that the proposed model significantly outperforms the baseline probabilistic DLR forecasting models regarding reliability and sharpness while using the fewest parameters.
\end{abstract}

\section{Introduction}\label{Sec:Intro}
Traditionally, transmission lines have been operated based on \jk{static line rating (SLR)}, which defines the maximum allowable current a transmission line can carry and remain constant over time. 
SLRs are calculated using conservative assumptions for weather conditions, such as high ambient temperatures and low wind speeds, to ensure the safe and reliable operation of power systems. 
However, these conservative assumptions often lead to underutilizing the additional available capacity of transmission lines \cite{fernandez2016review}.

To fully utilize the additional capacity of transmission lines, \jk{dynamic line rating (DLR) has} emerged as a promising solution \cite{douglass2019review}. \jk{DLR adjusts} the line ratings in real-time based on actual weather conditions, thereby increasing the overall power transfer capability of transmission lines. 
This approach is cost-effective as it increases transmission capacity without the installation of additional infrastructure.
However, employing accurate DLRs is challenging due to the inherent uncertainty of weather conditions, which hinders the integration of DLRs into grid operations. 
Therefore, developing accurate DLR forecasting models is of great interest \cite{gao2023day}.

\begin{figure}[t]
	\centering
\includegraphics[width=0.75\columnwidth]{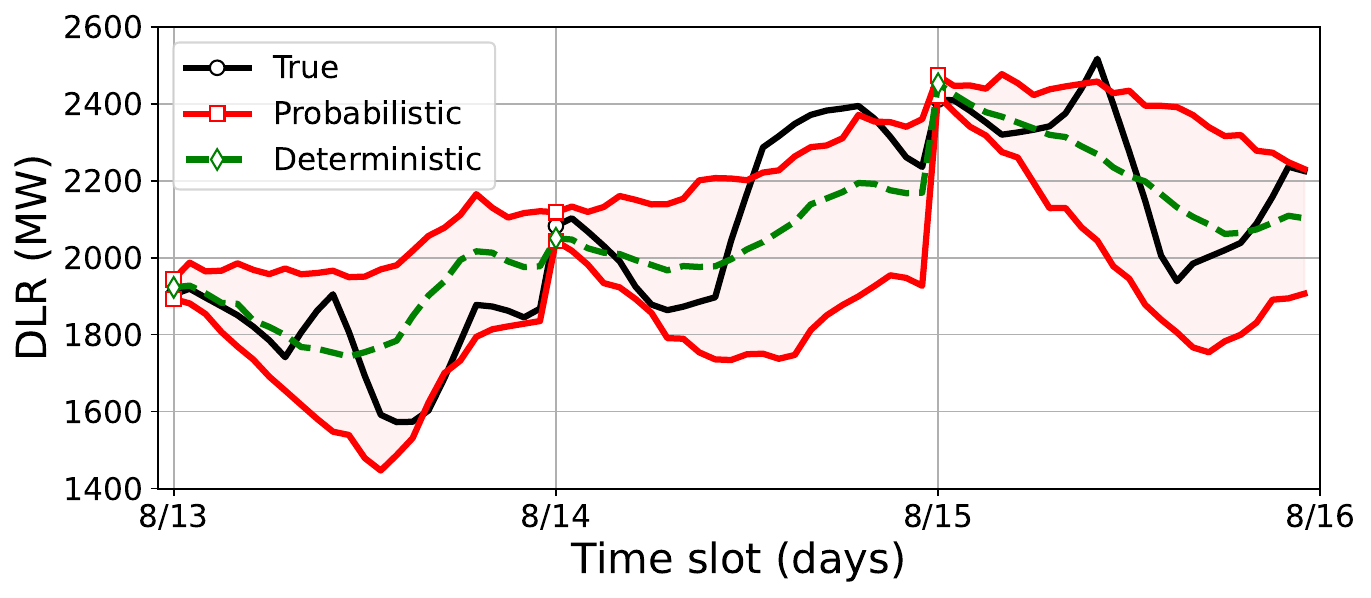}
\vspace{-7mm}
	\caption{\textcolor{black}{\small Probabilistic and deterministic DLR forecasting.}\vspace{-8mm}}
	\label{fig:ex_prob_deter}
\end{figure}

While there is a huge potential in DLR, several challenges exist. 
\textcolor{black}{\textbf{First}, deterministic forecasting inevitably contains forecasting errors, as illustrated in Fig.~\ref{fig:ex_prob_deter}, which can lead to the risk of either overloading or underutilizing the transmission line's capacity. }
Thus, probabilistic DLR forecasting to deal with uncertain weather conditions is essential. 
\textcolor{black}{\textbf{Second}, existing approaches focus on individual transmission lines without considering spatial correlations and interactions within the network.} 
However, incorporating these network-wide correlations is crucial for enhancing overall forecasting performance \cite{song2024graph}.

There have been several efforts in literature to resolve these two challenges. As regards the first challenge, quantile regression forests were employed for forecasting DLR in \cite{dupin2019optimal}. 
A Gaussian mixture model was used in \cite{viafora2020chance}, while \cite{madadi2019dynamic} utilized stochastic processes to model historical weather or DLR data for probabilistic forecasting.
However, these works do not fully address the second challenge, as they forecast ratings for only a limited number of lines without considering spatial correlations across the network. 
The authors of \cite{sun2022spatio} address both spatial and temporal correlation alongside probabilistic forecasting. 
\textcolor{black}{But the challenges still remain as} their approach considers only a limited subset of lines based on data from nearby weather stations,
and it does not capture the extended spatial correlations across the entire transmission network.

Recent advancements in graph convolutional networks (GCNs) offer promising tools to overcome the second challenge \cite{kipf2016semi}. 
By using message passing to aggregate information from neighboring nodes, GCNs can effectively learn the spatial correlation across the network. 
\textcolor{black}{The value of GCNs has been explored in various applications of power systems \cite{liao2021review}, but their application in DLR remains largely unexplored.}

In this regard, we propose a novel DLR forecasting algorithm to overcome the aforementioned two challenges. 
To deal with the first challenge, \textcolor{black}{our proposed method forecasts the prediction interval of uncertain DLRs based on quantile forecasting \cite{wang2019probabilistic}}. To address the second challenge, \textcolor{black}{the proposed method consists of a line graph convolutional network integrated with an LSTM to capture both spatial and temporal correlations across the transmission network.} We summarize our key contributions as follows:
\begin{enumerate}
    \item We propose a novel network-wide probabilistic DLR forecasting framework called double-hop line graph convolutional LSTM (D-LGCLSTM) that combines a double-hop line graph convolutional network with LSTM to effectively capture complex spatio-temporal patterns in transmission networks. \textcolor{black}{By utilizing double-hop message passing, D-LGCLSTM captures extended spatial correlations and reduces feature duplication within a single layer.} To the best of our knowledge, this is the first work that provides probabilistic DLR forecasting by incorporating both spatial and temporal information across entire transmission networks.
    \item We find that the forecasting performance of the single-hop line graph convolutional LSTM (hereafter referred to as LGCLSTM) is degraded due to feature duplication, where similar inputs are repeatedly aggregated during the message-passing process. We show that the proposed D-LGCLSTM can effectively mitigate the adversarial effect of feature duplication and capture extended spatial patterns across the network while using 65\% fewer parameters compared to LGCLSTM.
    \item We rigorously evaluate D-LGCLSTM against \textcolor{black}{four} state-of-the-art algorithms \textcolor{black}{ and LGCLSTM in probabilistic DLR forecasting \cite{meinshausen2006quantile, gao2023day, zhao2019t,simeunovic2021spatio}} on the Texas 123-bus backbone transmission system using five years of historical data. We extensively demonstrate that D-LGCLSTM outperforms all the baselines in terms of reliability, sharpness, and the number of learnable parameters.
\end{enumerate}

\section{Overall Framework and Methodology}\label{sec:FDIA}

\begin{figure}[t]
	\centering
\includegraphics[width=0.75\columnwidth]{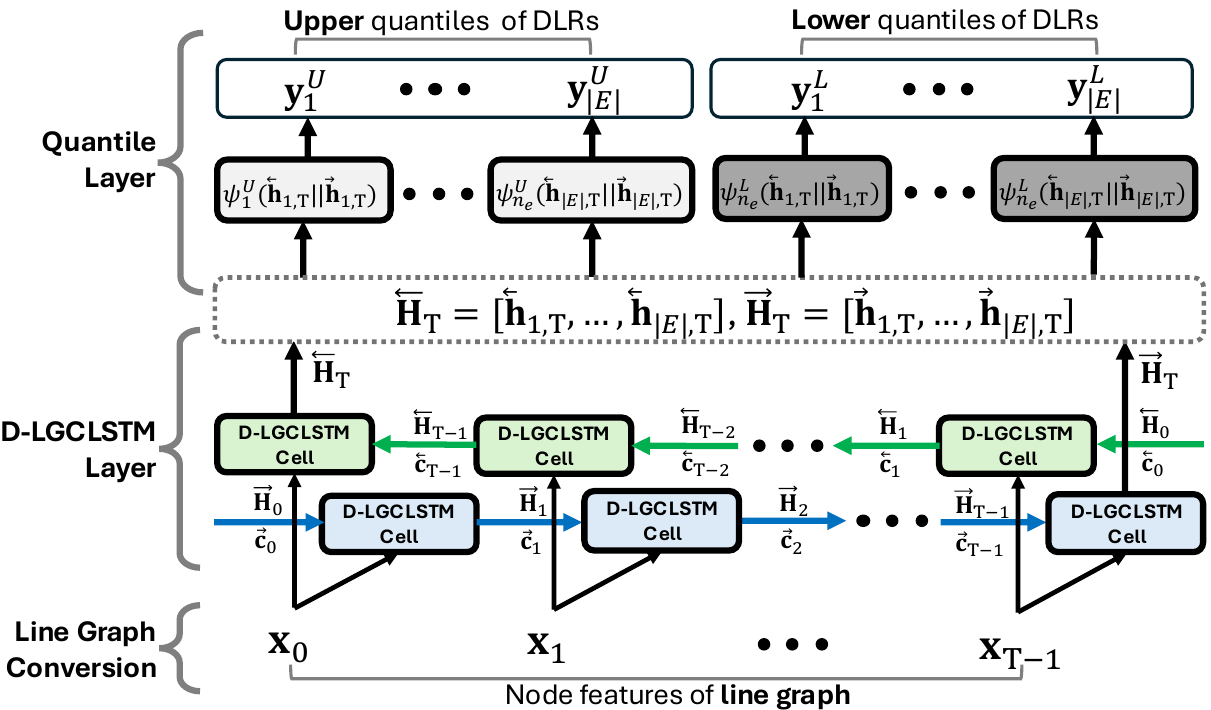}
\vspace{-4mm}
	\caption{\small Overall Framework of the proposed D-LGCLSTM.\vspace{-7mm}}
	\label{fig:dlr_overall_framework}
\end{figure}

As illustrated in Fig.~\ref{fig:dlr_overall_framework}, the overall framework of D-LGCLSTM includes a line graph conversion layer that transforms the transmission network into a line graph, a D-LGCLSTM layer that leverages both temporal and spatial features of the input data, and a quantile layer that produces probabilistic DLR forecasts for each line. From now, we will discuss the advantages and respective operations of each layer.

\subsection{Consistent node feature dimensions of a line graph.}
Let $G = (V, E)$ denote a graph where $V = \{v_1, ..., v_n\}$ is the set of nodes and $E\subseteq\{\{a, b\}|a,b\in V, a \neq b\}$ is the set of edges. Let $f_V:V\rightarrow\mathbb{R}^{n_v}$ and $f_E:E\rightarrow\mathbb{R}^{n_e}$ map nodes $v\in V$ and edges $e\in E$ to their feature vector where $n_v$ and $n_e$ are the dimensions of the feature, respectively.

The primary challenges arise from the need to integrate both node and edge features to apply GCN. However, GCN is inherently designed to operate on node features and does not directly utilize edge features \cite{kipf2016semi}. To integrate node and edge features into GCN, we concatenate the features of each edge onto its adjacent nodes. Let $R(v) = \{e\in E|v\in e\}$ be the set of edges incident to $v$. Then, we have
\begin{equation}
    \mathbf{x}_v = \Bigg(\concat_{e\in R(v)}f_E(e) \Bigg)\concat f_V(v),
\end{equation}
where $\mathbf{x}_v\in\mathbb{R}^{|R(v)|n_e + n_v}$ is the result of feature concatenation. Here, $\concat$ denotes vector concatenation. In a power network, $|R(v)|$ varies significantly. This variability leads to inconsistency of $\text{dim}(\mathbf{x}_v) = |R(v)|n_e + n_v$ across all $v\in V$. Furthermore, this is problematic for GCNs, which require a fixed feature dimension across all nodes for matrix multiplications and batch processing.

Alternatively, we concatenate the features of each node onto its connected edges. Let $S(e) = \{v\in V|v\in e\}$ be the set of nodes connected by $e\in E$. Then, we have \begin{equation}
    \mathbf{x}_e = \Bigg(\concat_{v\in S(e)}f_V(v) \Bigg)\concat f_E(e),
\end{equation}
where $\mathbf{x}_e\in\mathbb{R}^{|S(e)|n_v + n_e}$. Since each transmission line connects exactly two \textcolor{black}{buses} in power networks, $|S(e)| = 2$ for all $e\in E$. Thus, $\text{dim}(\mathbf{x}_e) = 2n_v+n_e$ is consistent for all edges. However, we cannot apply GCN to learn the concatenated edge features since it can only deal with nodes.

To leverage the consistency of edge feature dimensions, we employ the line graph convolutional network (LGCN) as follows: First, we convert the graph $G$ to its line graph $L(G) = (V_L, E_L)$ where each node $u\in V_L$ corresponds to an edge $e\in E$ from the original graph $G$ and $E_L = \{\{u_{e_i}, u_{e_j}\}|e_i, e_j \in E, e_i\neq e_j\}$. Thus, by using a line graph, we effectively treat each edge in $G$ as a node in $L(G)$.

\subsection{Reducing Feature Duplication via Double-Hop LGCN}

While LGCN successfully addresses the inconsistency of feature dimensions, it can suffer from feature duplication. Let $e_i = \{v_i, v_j\}\in E$, $e_j = \{v_j, v_k\}\in E$, and $e_k = \{v_k, v_l\}\in E$ denote edges in $G$, and share the node $v_{j}$ and $v_{k}$. Let $u_{e_i}$, $u_{e_j}$, and $u_{e_k}$ denote the nodes of $L(G)$ corresponding to these edges. Then, the feature vectors for these nodes are \textcolor{black}{defined as}\vspace{-4mm}
\begin{subequations}\label{eq:line_graph_feature}
\begin{align}
&\mathbf{x}_{u_i} = f_V(v_i)||f_V(v_j)||f_E(e_i),\\
&\mathbf{x}_{u_j} = f_V(v_j)||f_V(v_k)||f_E(e_j),\\
&\mathbf{x}_{u_k} = f_V(v_k)||f_V(v_l)||f_E(e_k).
\end{align}
\end{subequations}
Since both $\mathbf{x}_{u_i}$ and $\mathbf{x}_{u_j}$ contain the node feature $f_V(v_j)$, and both $\mathbf{x}_{u_j}$ and $\mathbf{x}_{u_k}$ contain the node feature $f_V(v_k)$, the features are duplicated when using LGCN \textcolor{black}{that aggregates features from single-hop neighbors}. This is problematic because using similar input features repeatedly may cause overfitting of the model and degrade its performance \cite{ying2019overview}. 

To mitigate the feature duplications, we \textcolor{black}{propose double-hop LGCN (D-LGCN) aggregates features from the double-hop neighbors. For example, when applying D-LGCN to $u_i$, it aggregates $\mathbf{x}_{u_i}$ and $\mathbf{x}_{u_j}$ in (\ref{eq:line_graph_feature}). Interestingly, D-LGCN effectively skips single-hop neighbors and avoids feature duplication since $\mathbf{x}_{u_i}$ and $\mathbf{x}_{u_j}$ does not share any node features of original graph $G$, which is different from LGCN.}

\textcolor{black}{Another significant benefit of using D-LGCN is that it requires a lower number of learnable parameters compared to LGCN to aggregate the features from multiple-hop neighbors. LGCN requires stacking multiple graph convolution layers to aggregate features from multiple-hop neighbors. Thus, LGCN needs $k$ layers to aggregate features from $k$-hop neighbors. By contrast, D-LGCN only requires $k/2$ graph convolution layers since it can aggregate features from double-hop neighbors. Note that although the number of learnable parameters for D-LGCN is at least twice less than LGCN, it shows superior performance compared to LGCN and other baselines. We will discuss this in Section~\ref{sec:results}.}

\subsection{Embedding Double-Hop LGCNs into LSTM}\label{subsec:prediction}
Now, we propose D-LGCLSTM by embedding D-LGCN into LSTM. \textcolor{black}{Let $\tilde{\mathbf{A}} = \mathbf{A}_d + \mathbf{I}$ where $\mathbf{A}_d$ is adjacency matrix of double-hop neighbors and $\mathbf{I}$ is identity matrix \cite{kipf2016semi}. Let $\mathbf{x}_{u_i,t}$ and $\mathbf{h}_{i,t}$ denote the feature and hidden vector of $i$th node of line graph $L(G)$ at time slot $t$, respectively. Then, we have the matrix of feature vectors $\mathbf{X}_{t} = [\mathbf{x}_{u_1, t}, ..., \mathbf{x}_{u_{|E|}, t}]$ and hidden vectors $\mathbf{H}_{t} = [\mathbf{h}_{1,t}, ..., \mathbf{h}_{|E|,t}]$.} D-LGCLSTM cell at time slot $t$ consists of the forget gate $\mathbf{f}_t$, the input gate $\mathbf{i}_t$, the output gate $\mathbf{o}_t$ and the candidate cell state gate $\mathbf{g}_t$. Then, the hidden state $\mathbf{H}_t$ and cell state $\mathbf{c}_t$ are updated as follows:
\begin{subequations}
	\begin{align}
		&\mathbf{f}_t = \sigma(\tilde{\mathbf{A}}\mathbf{X}_{t-1}\mathbf{W}_f + \mathbf{H}_{t-1}\mathbf{U}_f + \mathbf{b}_f),\\
		&\mathbf{i}_t = \sigma(\tilde{\mathbf{A}}\mathbf{X}_{t-1}\mathbf{W}_i + \mathbf{H}_{t-1}\mathbf{U}_i + \mathbf{b}_i),\\
		&\mathbf{o}_t = \sigma(\tilde{\mathbf{A}}\mathbf{X}_{t-1}\mathbf{W}_o + \mathbf{H}_{t-1}\mathbf{U}_o + \mathbf{b}_o),\\
		&\mathbf{g}_t = \tanh(\tilde{\mathbf{A}}\mathbf{X}_{t-1}\mathbf{W}_g + \mathbf{H}_{t-1}\mathbf{U}_g + \mathbf{b}_g),\\
		&\mathbf{c}_t = \mathbf{f}_t \odot \mathbf{c}_{t-1} + \mathbf{i}_t \odot \mathbf{g}_t,\\
		&\mathbf{H}_t = \mathbf{o}_t \odot \sigma(\mathbf{c}_t),
	\end{align}\label{eq:D-LGCN}
\end{subequations}\vspace{-5mm}

\noindent where $\mathbf{W}_f$, $\mathbf{W}_i$, $\mathbf{W}_o$, and $\mathbf{W}_g$ are learnable weight matrices associated with the input features, $\mathbf{U}_f$, $\mathbf{U}_i$, $\mathbf{U}_o$, and $\mathbf{U}_g$ are learnable weight matrices of hidden state, and $\mathbf{b}_f$, $\mathbf{b}_i$, $\mathbf{b}_o$, and $\mathbf{b}_g$ are learnable biases. $\sigma$ is the sigmoid function and $\odot$ is the element-wise product. Note that we only substitute the input sequence part of LSTM and left the hidden state part as it was to avoid oversmoothing due to repeatedly applying graph deep learning to hidden vectors \cite{chen2020measuring}. \textcolor{black}{Additionally, we use a bidirectional approach to capture spatial and temporal patterns from both the past and present. Thus, $\overrightarrow{\mathbf{H}}_t$ and $\overleftarrow{\mathbf{H}}_t$ in Fig.~\ref{fig:dlr_overall_framework} denote the hidden matrices of the forward and backward D-LGCLSTM cell at time slot $t$.}

\subsection{Quantile Layer for Probabilistic Forecasting}
\textcolor{black}{For probabilistic forecasting, we use $\psi_i^U$ and $\psi_i^L$, which are two layers of neural networks and map the output of D-LGCLSTM layer in Fig.~\ref{fig:dlr_overall_framework} to the prediction intervals of each line, where $i\in\{1,...,|E|\}$. Specifically, let $\mathbf{y}_i^U = \psi_i^U(\overleftarrow{\mathbf{h}}_{i,T}||\overrightarrow{\mathbf{h}}_{i,T})$ and $\mathbf{y}_i^L = \psi_i^L(\overleftarrow{\mathbf{h}}_{i,T}||\overrightarrow{\mathbf{h}}_{i,T})$ denote the upper and lower quantile forecasts of the next day's DLR of $i$th line. $\overleftarrow{\mathbf{h}}_{i,T}$ and $\overrightarrow{\mathbf{h}}_{i,T}$ are the hidden states from the backward and forward hidden matrices $\overleftarrow{\mathbf{H}}_T$ and $\overrightarrow{\mathbf{H}}_T$ at final time slot $T$, respectively. These estimated quantiles serve as the lower and upper bounds of the prediction interval. Now, let $q \in \{L, U\}$ represent the lower and upper quantiles, and let $Q_q$ be the corresponding quantile levels. Then, the quantile loss function for the $i$th line is defined as \cite{wang2019probabilistic}
\begin{equation}
	\mathcal{L}(y_{i,t}^q, y_{i,t}) = \begin{cases}
		Q_q(y_{i,t} - y_{i,t}^q), \;\;\quad\quad\quad y_{i,t}^q\leq y_{i,t},\\
		(1-Q_q)(y_{i,t}^q - y_{i,t}), \quad otherwise,
	\end{cases}\label{eq:quantile_loss}
\end{equation}
where $y_{i,t}^q$ and $y_{i,t}$ are $t$th element of $\mathbf{y}_i^q$ and true DLR $\mathbf{y}_i$, respectively. Finally, we use $\sum_{q \in \{L, U\}}\sum_{i=1}^{|E|}\sum_{t=1}^{\tau}\mathcal{L}(y_{i,t}^q, y_{i,t})$ to train the model for all $q$ where $\tau$ is prediction horizon.}

\section{\textcolor{black}{Case Studies}}
\subsection{\textcolor{black}{Simulation Settings}}
\subsubsection{Data Preparation}
We use the Texas 123-bus backbone transmission (TX-123BT) system \cite{lu2023synthetic} to verify the performance of the proposed method in probabilistic DLR forecasting. This system contains 123 buses and 244 lines. \textcolor{black}{For our experiments, we reduce the lines to 173 by retaining only one representative among parallel lines that have identical parameters (e.g., endpoints, weather, and rating data) to avoid redundant inputs and reducing computational overhead.} We utilize five years of historical weather data for each bus and DLR data for each line from January 1, 2017, to December 31, 2021, with a one-hour resolution. The weather data include measurements of temperature, wind speed, wind direction, and solar radiation. The DLR data consist of line ratings calculated based on the heat balance equation \textcolor{black}{based on IEEE 738 standard} \cite{IEEE738_2012} \textcolor{black}{as done in \cite{madadi2019dynamic}}. We split the dataset into a training set and a testing set using a 4:1 ratio, \textcolor{black}{where the 2021 test data notably encompasses the 2021 Texas power crisis.} Each bus includes the previous seven days of historical weather data and its geographical coordinates. Each line includes the previous seven days of historical DLR data, its length, and the current season (spring, summer, fall, or winter). The model uses these input data to forecast the next day's DLR for each line. \textcolor{black}{Note that all the experiments are conducted on NVIDIA A100 SXM4 40 GB GPU and Intel Xeon 2.20 GHz CPU.}

\begin{figure}[t]
	\centering
\includegraphics[width=0.7\columnwidth]{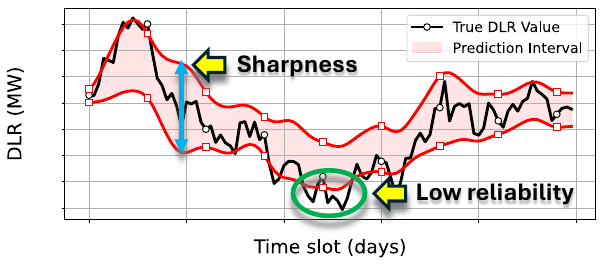}
\vspace{-3.5mm}
	\caption{\textcolor{black}{\small An example of reliability and sharpness.}\vspace{-3.5mm}}
	\label{fig:eval_metric}
\end{figure}

\begin{table}[t]
	\centering
    \captionsetup{justification=centering, labelsep=period, font=footnotesize, textfont=sc}
	\caption{\textcolor{black}{Comparison of the baselines methods. $\dag$ represents the proposed method.} \vspace{-2mm}}
	\label{table:baseline_methods}
	\begin{tabular}{c|cccc}
		\toprule
		\textbf{Method} & \textbf{Scale} & \textbf{\makecell{Line\\Graph}} & \textbf{\makecell{Hop\\Count}} & \textbf{\makecell{The Num.\\of Layers}} \\
		\midrule\midrule
        \textcolor{black}{QRF \cite{meinshausen2006quantile}} &  \textcolor{black}{Single line} &  \textcolor{black}{$\times$} &  \textcolor{black}{--} &  \textcolor{black}{--} \\
		LSTM \cite{gao2023day} & Single line & $\times$ & -- & 1 \\
		T-GCN \cite{zhao2019t} & Network & $\times$ & Single-hop & 3 \\
		GCLSTM \cite{simeunovic2021spatio} & Network & $\times$ & Single-hop & 3 \\
        \midrule
		$\text{LGCLSTM}^\dag$ & Network & \checkmark & Single-hop & 2 \\
		$\text{D-LGCLSTM}^\dag$ & Network & \checkmark & Double-hop & 1 \\
		\bottomrule
	\end{tabular}\vspace{-7mm}
\end{table}

\subsubsection{Evaluation Metrics}

\textcolor{black}{We use four evaluation metrics to reflect reliability and sharpness, which are illustrated in Fig.~\ref{fig:eval_metric}. In probabilistic forecasting, reliability refers to how well the prediction intervals capture the actual DLR values; low reliable prediction intervals may lead to overheating or underutilization of transmission lines. Sharpness indicates the narrowness of the prediction intervals; sharper intervals enable operators to maximize line utilization. We measure reliability with the average coverage error (ACE) and sharpness with the prediction interval normalized average width (PINAW) \cite{li2022integrated}. We also use the interval score (IS) and quantile score (QS) \cite{pinson2006properties} to evaluate both aspects since sharper intervals are desirable when reliability is maintained. The detailed mathematical definitions of the metrics are provided in\cite{li2022integrated,pinson2006properties}.}

\begin{table}[t]
	\centering
    \captionsetup{justification=centering, labelsep=period, font=footnotesize, textfont=sc}
	\caption{Performance comparisons of probabilistic DLR forecasting models. The best results are in \textbf{bold}.\vspace{-2mm}}
	\label{table:simulation_results}
	\begin{tabular}{c|ccccc}
		\toprule
		\textbf{Method} & \textbf{\makecell{ACE\\(\%)}} & \textbf{\makecell{PINAW\\(\%)}} & \textbf{\makecell{IS\\(\%)}} & \textbf{\makecell{QS\\(\%)}} & \textbf{\makecell{The Num.\\of Params.\\ ($\times10^7$)}} \\
		\midrule\midrule
        \textcolor{black}{QRF \cite{meinshausen2006quantile}}                     & \textcolor{black}{\textbf{1.31}} & \textcolor{black}{47.00} & \textcolor{black}{18.22} & \textcolor{black}{2.73} & \textcolor{black}{--}  \\
		LSTM \cite{gao2023day}                     & 5.40 & 36.57 & 13.50 & 2.03 & 1.55   \\
		T-GCN \cite{zhao2019t}                   & 3.41 & 42.35 & 13.19 & 1.97 & 99.84  \\
		GCLSTM \cite{simeunovic2021spatio}                 & 4.60 & 37.90 & 13.56 & 2.05 & 7.02   \\
        \midrule
		$\text{LGCLSTM}^\dag$                   & 2.87 & 38.62 & 13.17 & 2.01 & 4.25   \\
		$\text{D-LGCLSTM}^\dag$      & 2.74 & \textbf{34.91} & \textbf{12.66} & \textbf{1.91} & \textbf{1.42}   \\
		\bottomrule
	\end{tabular}\vspace{-7mm}
\end{table}

\subsubsection{Baseline Models}
We compare the proposed D-LGCLSTM against five baselines \textcolor{black}{in Table~\ref{table:baseline_methods}}. LSTM \cite{gao2023day} captures only temporal patterns of a single line. T-GCN \cite{zhao2019t} combines GCN and LSTM sequentially but does not integrate the GCN into the LSTM cell. In contrast, GCLSTM \cite{simeunovic2021spatio} integrates the GCN directly into the LSTM cell. Both T-GCN and GCLSTM operate on the original graph without transforming it into a line graph. LGCLSTM applies GCLSTM after line graph conversion for consistent feature dimension. D-LGCLSTM advances further by aggregating the features over double-hop neighbors in the line graph. \textcolor{black}{Note that since D-LGCLSTM leverages a single layer of double-hop LGCN, it requires fewer parameters and offers faster inference compared to GCLSTM and LGCLSTM, which rely on multiple layers.} \textcolor{black}{We also include Quantile Regression Forests (QRF) \cite{meinshausen2006quantile} as a non-deep-learning reference method that forecasts single-line ratings without leveraging graph structure.} \textcolor{black}{We train LSTM, GCLSTM, LGCLSTM, and D-LGCLSTM for 50 epochs, and T-GCN for 20 epochs using AdamW (learning rate and weight decay of $10^{-3}$). All models use an LSTM hidden dimension of 64, with a mini-batch size of 64 for T-GCN, GCLSTM, and LGCLSTM, and 128 for D-LGCLSTM.}

\subsection{\textcolor{black}{Results}}\label{sec:results}
\subsubsection{Overall Performance Comparisons}\label{sec:overall}
Table~\ref{table:simulation_results} provides a comparison of DLR forecasting performance across the baselines. The proposed D-LGCLSTM outperforms baseline models in all evaluation metrics. Specifically, D-LGCLSTM achieves nearly half the ACE of LSTM. In addition, LSTM shows a higher ACE compared to all the models in Table~\ref{table:simulation_results}, demonstrating the necessity to use a graph-based model for reliable forecasting. \textcolor{black}{Although QRF achieves the lowest ACE among all methods, it exhibits the worst PINAW. By contrast,} D-LGCLSTM achieves a significantly lower PINAW and thus obtains sharper prediction intervals compared to T-GCN, which does not integrate GCN or LGCN into LSTM. 

Moreover, D-LGCLSTM outperforms all the baselines in IS and QS, which measure both reliability and sharpness. \textcolor{black}{By contrast, QRF achieves the worst IS and QS scores due to its unbalanced performance, which implies that simply minimizing ACE does not guarantee practical intervals.} D-LGCLSTM reduces nearly 7\% of IS and QS from GCLSTM by applying a double-hop message passing in a line graph. \textcolor{black}{Thus, D-LGCLSTM successfully achieves high reliability while keeping the prediction intervals as sharp as possible. This is highly beneficial from a power system perspective, as it enables operators to make more informed and precise decisions by maximizing transmission line utilization without unnecessary conservatism.}

\begin{figure}[t]
	\centering
\includegraphics[width=0.75\columnwidth]{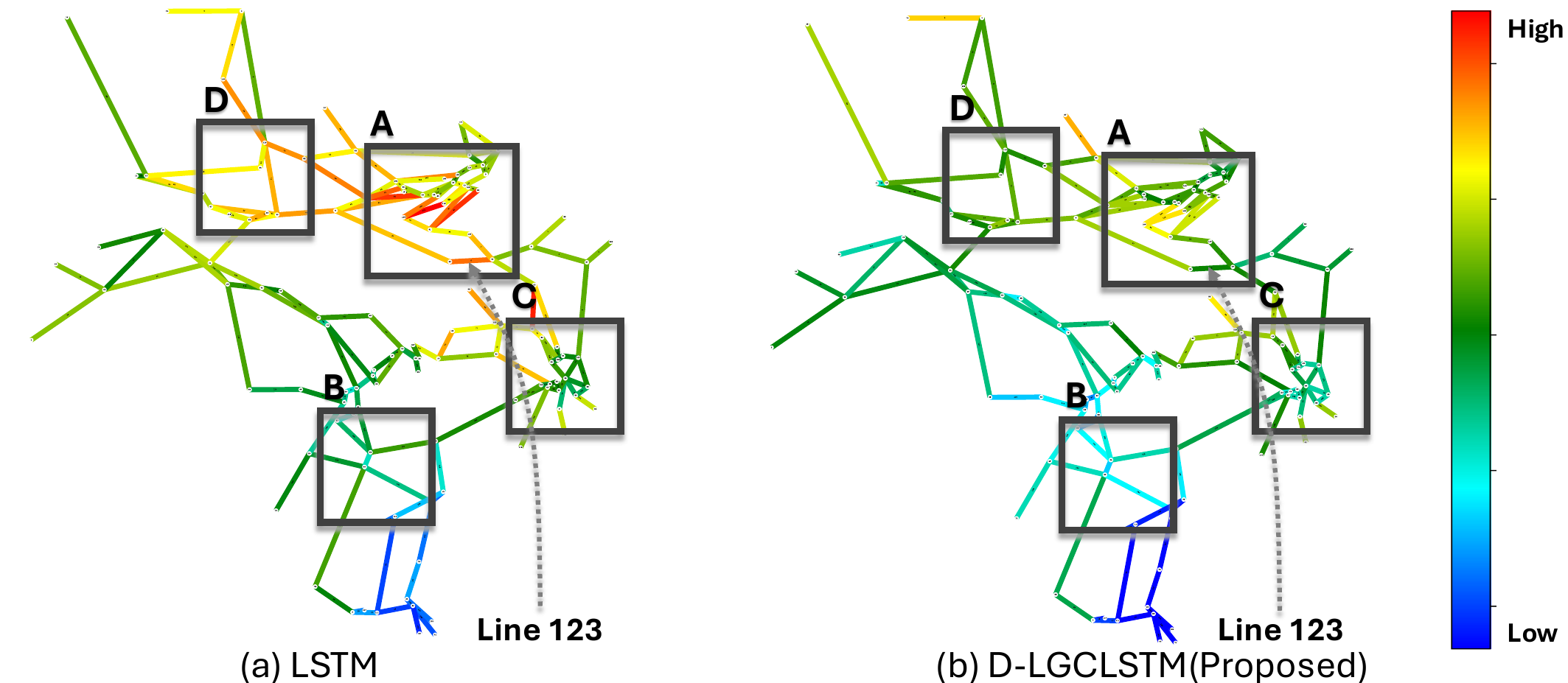}
\vspace{-2mm}
	\caption{\small  Heat maps of
the average QS for the lines in TX-123BT.}
	\label{fig:heat_map_QS}\vspace{-5mm}
\end{figure}

\textcolor{black}{In addition to the significant forecasting performance and benefits for power systems, D-LGCLSTM reduces the number of parameters by approximately 80\% and 99\% compared to GCLSTM and T-GCN.} This is due to the double-hop message passing on the line graph, which captures extended spatial patterns with fewer layers. Notably, although T-GCN has the highest number of parameters among the models, it does not achieve the best results. This indicates that increasing model complexity does not necessarily improve the performance.

\begin{figure}[t]
     \centering
     \begin{subfigure}[b]{0.8\columnwidth}
     \centering
         \includegraphics[width=\columnwidth]{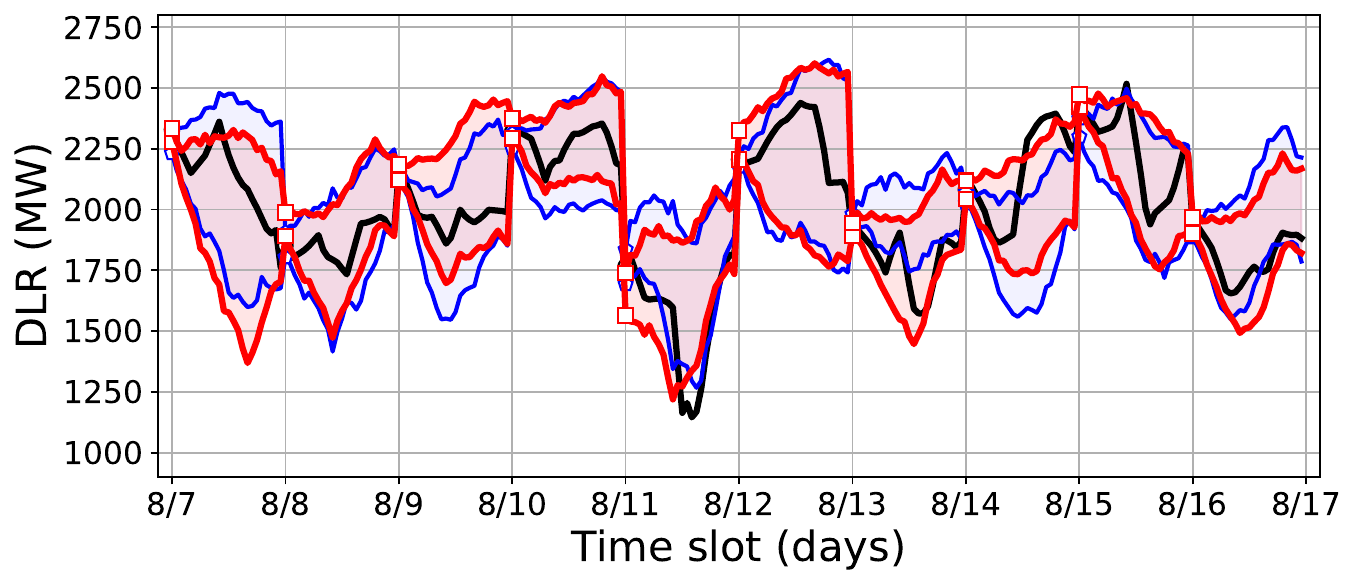}
         \vspace{-7mm}
         \caption{\small Probabilistic DLR forecasting (qunatile: 0.1$-$0.9).}
         \label{fig:prob_forecasting}
     \end{subfigure}
     \begin{subfigure}[b]{0.8\columnwidth}
     \centering
         \includegraphics[width=\columnwidth]{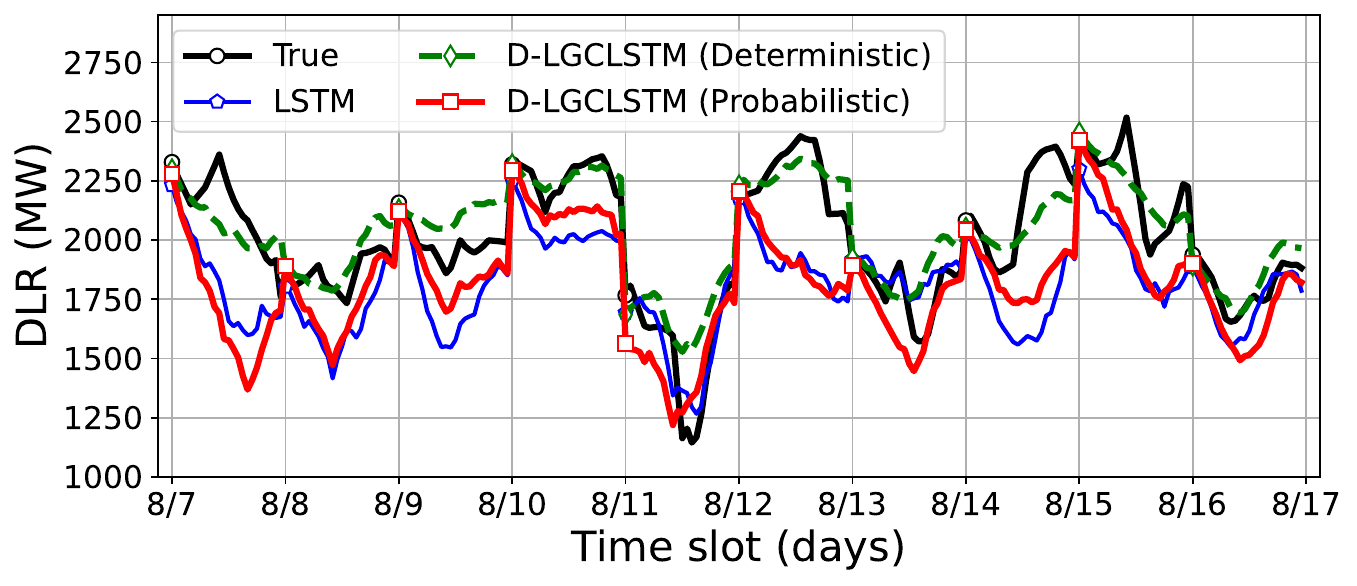}
         \vspace{-8mm}
         \caption{\small Robust \textcolor{black}{(lowest value)} DLR forecasting.  \vspace{-5mm}}
         \label{fig:robust_forecasting}
     \end{subfigure}
     \caption{\small Probabilistic and robust DLR forecasting for line 123. \vspace{-7mm}}
     \label{fig:forecastings}
\end{figure}

\subsubsection{Benefits of Network-Wide Consideration}
To demonstrate the benefits of incorporating spatial features in probabilistic DLR forecasting, we illustrate heat maps of the average QS for each transmission line using test data across the test system in Fig.~\ref{fig:heat_map_QS}. Specifically, we compare the performance of LSTM and D-LGCLSTM to verify the importance of spatial information for accurate probabilistic forecasting. 

In Fig.~\ref{fig:heat_map_QS}, \textcolor{red}{red} indicates high QS (poorer performance), while \textcolor{blue}{blue} represents low QS (better performance). As can be seen, D-LGCLSTM generally exhibits lower QS across the network compared to LSTM. In particular, D-LGCLSTM achieves significant improvements in QS in regions A, B, C, and D where neighboring buses are densely clustered. The improvements in these regions indicate the existence of a strong spatial correlation among transmission lines that can be effectively captured by the D-LGCLSTM. Unlike LSTM which treats each line independently and only captures temporal patterns, D-LGCLSTM leverages both temporal features and the network topology through line graph and double-hop message passing. By doing so, D-LGCLSTM successfully produces more accurate and reliable DLR forecasting.

\subsubsection{Robust DLR Forecasting}

Now, we focus on a specific transmission line to compare the performance of LSTM and D-LGCLSTM, and their applicability to grid operations through robust DLR forecasting as shown in Fig.~\ref{fig:forecastings}. Specifically, we select line 123, which exhibits the largest improvement in QS when transitioning from LSTM to D-LGCLSTM. We analyze 10 days during the summer when ambient temperatures are high and transmission lines are more susceptible to overheating. Fig.~\ref{fig:prob_forecasting} presents the DLR forecasting results for line 123 using both LSTM (blue line) and D-LGCLSTM (red line). While the prediction intervals generated by both methods generally capture the actual DLR values (black line), the prediction interval of LSTM fails to encompass the actual DLR values from August 13 to August 15, whereas D-LGCLSTM successfully captures them.

To evaluate the applicability of the DLR forecasts in grid operations, we employ the lower bound of the prediction intervals as robust DLR forecasts to prevent unexpected overheating while utilizing the additional available capacity of the line. As illustrated in Fig.~\ref{fig:robust_forecasting}, we also include deterministic DLR forecasts (green dashed line) that do not consider uncertainty. Although deterministic forecasting captures the overall trends of the true DLR, it inevitably contains forecasting errors that could risk overloading or underutilization of the transmission line's capacity. In contrast, the robust forecasts derived from both LSTM and D-LGCLSTM are generally lower than the true DLR values, providing a safety margin against overloading. However, the robust forecasts from LSTM are relatively conservative (e.g., during August 9–11 and August 14–15) and less reliable (e.g., during August 13–14) compared to those from D-LGCLSTM which is more suitable for grid operations.

\section{Conclusion}

In this paper, we proposed a novel network-wide probabilistic dynamic line rating (DLR) forecasting model called double-hop line graph convolutional LSTM (D-LGCLSTM), which integrates line graph convolutional networks into LSTM to incorporate both spatial and temporal information. By employing a double-hop message passing on the line graph, D-LGCLSTM captures extended spatial correlations and mitigates feature duplication in single-hop models. The simulations on the Texas 123-bus backbone transmission system demonstrate that D-LGCLSTM outperforms all the baselines in terms of reliability and sharpness while using the least number of parameters. Specifically, D-LGCLSTM achieves up to a 7\% improvement in IS and QS and reduces the number of model parameters by at most 99\% compared to baselines. 

\textcolor{black}{Although D-LGCLSTM effectively captures true DLR across the network better than the baselines, abrupt or extreme changes in DLR remain challenging.} For future work, we plan to integrate D-LGCLSTM with grid operations, such as security-constrained unit commitment or market operations, \textcolor{black}{including applications to larger or more diverse power networks, and investigate how operators can select suitable quantiles for real-world operational needs.} \textcolor{black}{We also plan to conduct a more detailed investigation of the model’s performance under extreme weather events to ensure reliable DLR forecasting.}

\balance
\bibliographystyle{IEEEtran}
\bibliography{lgcn_dlr.bib}
\end{document}